# Shaping Magnetite Nanoparticles from First Principles

Hongsheng Liu and Cristiana Di Valentin*

*Dipartimento di Scienza dei Materiali, Università di Milano-Bicocca, via R. Cozzi 55,*

*I-20125 Milano, Italy*

*e-mail: cristiana.divalentin@unimib.it*

**Abstract**

Iron oxide magnetic nanoparticles (NPs) are stimuli-responsive materials at the forefront of nanomedicine. Their realistic finite temperature simulations are a formidable challenge for first-principles methods. Here, we use density functional tight binding to open up the required time and length scales and obtain global minimum structures of $Fe_3O_4$ NPs of realistic size (1400 atoms, 2.5 nm) and of different shapes, which we then refine with hybrid density functional theory methods to accomplish proper electronic and magnetic properties, which have never been accurately described in simulations. On this basis, we develop a general empirical formula and prove its predictive power for the evaluation of the total magnetic moment of $Fe_3O_4$ NPs. By converting the total magnetic moment into the macroscopic saturation magnetization, we rationalize the experimentally observed dependence with shape. We also reveal interesting reconstruction mechanisms and unexpected patterns of charge ordering.

Magnetite ($Fe_3O_4$) nanoparticles (NPs) are top-class materials for biomedical applications because of their excellent soft magnetism (high saturation magnetization and low coercive force), good biocompatibility, and low cytotoxicity [1,2]. They constitute the new generation contrast agents for magnetic resonance imaging (MRI) and are effective carriers for targeted drug delivery, heating agents in magnetic hyperthermia, adsorbents for magnetic bioseparation, and biosensors [3,4,5,6,7,8,9].

Monodisperse $Fe_3O_4$ NPs with sizes variable from 3 to 22 nm in diameter have been successfully prepared [10,11,12,13,14] with different shapes including cubes, octahedra, rhombic dodecahedra, truncated octahedra and spheres [15,16,17,18,19,20,21,22,23,24]. Magnetite NPs are found to be superparamagnetic above the blocking temperature and ferrimagnetic below it [11,12,14,15,16,23,24]. Furthermore, tunneling microscopy shows that below the Verwey temperature magnetite NPs are semiconductors with a small band gap from 0.14 to 0.30 eV [25,26].

Despite the relevance of magnetite NPs in nanobiotechnology, we observe a severe lack of a theoretical framework, which could assist in the interpretation of experimental findings at an atomic scale and guide further experiments. For instance, only recently the $\sqrt{2} \times \sqrt{2}$ reconstruction of the clean $Fe_3O_4$(001) single crystal surface was revealed by Bliem et al. [27] through a combined experimental and theoretical study. This begs the next question: what kind of reconstructions may arise in a nano-confined magnetite particle?

Unfortunately, magnetite is a complex material to be described accurately by theoretical methods. We have shown that, to catch proper structural, electronic and magnetic properties of even the most simple bulk and flat surface systems, high-level quantum mechanical (QM) techniques, beyond standard density functional theory (DFT), are required [28,29]. Up to now, magnetite nanoparticles have only been addressed by force-field methods to study their interaction with surfactants [30,31,32,33]. However, these types of simulations have some intrinsic limitations, since they cannot provide any information on the electronic and magnetic structure, cannot handle bond breaking and bond formation, and have limited transferability.

With the present Letter, we make a major breakthrough in the theoretical modelling of magnetite nanosystems. First, we solve the critical problem of the correct assignment of the total magnetic moment ($m_{tot}$) to magnetite model nanostructures and provide the community with a validated general empirical formula for its a-priori evaluation. From that we derive the saturation magnetization ($M_S$) of an ideal macroscopic sample of all identical NPs for comparison with experiments. Then, by combining density functional tight binding (DFTB) [34] and hybrid DFT methods [35], we accomplish the quantum mechanical simulation of $Fe_3O_4$ NPs of realistic size in both cubic (1466 atoms, edge length of 2.3 nm) and spherical (1006 atoms, diameter of 2.5 nm) shapes. Global minimum atomic-scale structures of the NPs are obtained by high-temperature annealing simulations with the Hubbard corrected DFTB (DFTB+U) method, followed by full atomic relaxation with hybrid DFT. Interesting reconstruction mechanisms and unexpected patterns of charge ordering are revealed. A rational basis for the larger experimentally-observed $M_S$ of nanocubes with respect to nanospheres is also derived from our results.

This Letter fills the existing gap in the quantum chemical description of magnetite NPs of realistic size and paves the way for further theoretical studies for the benefit of both the computational and experimental communities.

Here, molecular dynamics (MD) simulations were performed with self-consistent charge DFTB method, as implemented in the DFTB+ package [36], to search for the global minimum structure of $Fe_3O_4$ NPs. Then, hybrid functional calculations (the Heyd-Scuseria-Ernzerhof (HSE) screened hybrid functional [35]) were carried out using the CRYSTAL17 package [37,38] to get the final optimized structures and electronic and magnetic properties of the nanoparticles. Further computational details can be found in the Supplemental Material.

Considering the frequent observation of cubic $Fe_3O_4$ NPs enclosed by six (001) facets in experiments [15,16,17], we carved from the bulk a magnetite nanocube ($Fe_{602}O_{864}$) of 1466 atoms with edge length of 2.3 nm in a way that the lowest coordination of Fe and O atoms is four- and two-fold, respectively. We define two kinds of Fe ions in the NP, $Fe_{Tet}$ and $Fe_{Oct}$, depending on the occupied bulk lattice site (tetrahedral and octahedral), even when they become undercoordinated at the surface. The nanocube is in the $T_d$ symmetry and presents two types of corner sites (see Fig. 1): (a) four corners expose an $Fe_{Tet}$ ion at the apex (type I) and (b) four corners expose an O atom at the apex (type II).

The first challenge one must face to perform the QM simulation of a magnetite NP is the definition of its optimal total magnetic moment ($m_{tot}$), as discussed in detail in the Supplemental Material. To overcome this challenge, we propose (and validate in this work below) an empirical formula for $m_{tot}$,

$$m_{tot} = 5 \times \left(N(Fe^{3+}_{Oct}) - N(Fe^{3+}_{Tet})\right) + 4 \times \left(N(Fe^{2+}_{Oct}) - N(Fe^{2+}_{Tet})\right) \quad (1),$$

where $Fe_{Oct}^{3+}$ and $Fe_{Oct}^{2+}$ are $Fe^{3+}$ and $Fe^{2+}$ ions at octahedral sites, $Fe_{Tet}^{3+}$ and $Fe_{Tet}^{2+}$ are $Fe^{3+}$ and $Fe^{2+}$ ions at tetrahedral sites and N is the number of the corresponding ions. This formula can be interpreted by means of the crystal field theory and the $d$ orbitals occupation of different Fe ions in bulk magnetite, in line with what was previously observed by hybrid DFT calculations [28]: for $Fe_{Oct}^{3+}$ and $Fe_{Tet}^{3+}$, the high-spin $3d^5$ electron configuration gives an atomic magnetic moment of +5 and −5 $\mu_B$, respectively; for $Fe_{Oct}^{2+}$ and $Fe_{Tet}^{2+}$, the high-spin $3d^6$ electron configuration gives +4 $\mu_B$ and -4 $\mu_B$, respectively. For the nanocube, since all $Fe_{Tet}$ ions are assumed to be charged 3+ and O ions −2, $N(Fe_{Oct}^{3+})$ and $N(Fe_{Oct}^{2+})$ can be easily calculated. This formula works perfectly for magnetite bulk and (001) surfaces and gives $m_{tot} = 1232$ $\mu_B$ for the carved nanocube.

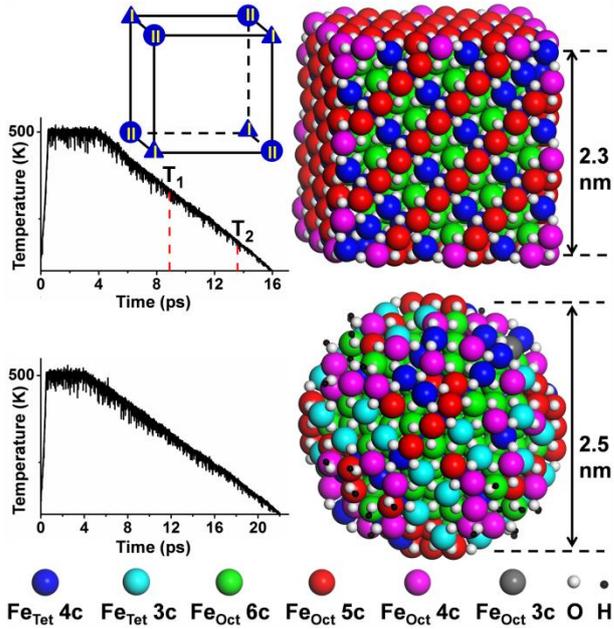

**Figure 1.** Simulated annealing temperature profiles and global minimum structures of the magnetite nanocube (up) and nanosphere (down). The color coding of atoms is given in the legend at the bottom. Labels 3c−6c labels indicate the actual coordination number of the corresponding ions. (Inset) Shows the two types of corners.

The nanocube global minimum structure was searched by MD simulations with the DFTB+U method that simulates a temperature annealing process up to 500 K (Fig. 1). We recently proved that DFTB+U, with our newly proposed parametrization of the Fe–O interactions, is very efficient and satisfactorily reliable for the description of bulk and surface magnetite [39]. With this computationally cheaper method, one can perform simulations for systems of a thousand atoms or more on timescales of tenths of picoseconds, which is well beyond what is currently accessible with first-principles MD simulations. During the annealing, the four type-I corners underwent reconstructions at time $T_1$ and $T_2$, as shown in Fig. 1. Compared with the structure before annealing, the reconstructed nanocube is about 14 meV per atom lower in energy. To clarify the reconstruction mechanism, the top and side views of one of the three (001) facets that meet at the type-I corner are displayed in Fig. 2. The process consists of the transfer of 3 six-coordinated $Fe_{Oct}$ ions around the corner (green large spheres) to 3 four-coordinated $Fe_{Tet}$ ions (marked as 1, 2 and 3). The top and side views of the other two (001) facets that meet at this corner are exactly the same as the one shown in Fig. 2, but with $Fe_{Tet}$ 2 or 3 in the place of $Fe_{Tet}$ 1, respectively. This reconstruction presents some analogies to that proposed for single crystal (001) surfaces [27]. However, in that case, every two Fe vacancies at octahedral sites in the subsurface layer are replaced only by one additional tetrahedral Fe, leading to an unbalanced Fe:O stoichiometry.

The DFTB+U optimized reconstructed nanocube structure was confirmed by full atomic relaxation with the more sophisticated HSE hybrid functional. For nanocubes of larger size, we expect the same reconstruction at the corners since this atomic rearrangement only involves few atoms around corners.

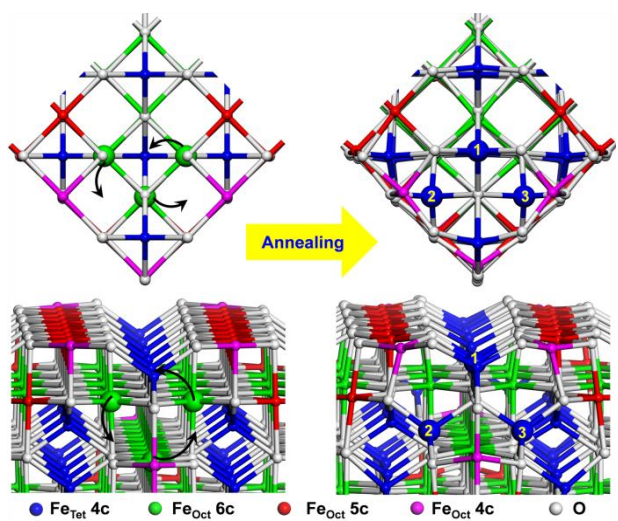

**Figure 2.** Top and side views of one of the three (001) facets that meet at the type-I corners of the nanocube, before and after annealing. The color coding of atoms is given in the legend at the bottom. Labels 3c−6c indicate the actual coordination number of the corresponding ions. The black arrows indicate the motion of atoms during the annealing process.

Note that, after reconstruction, the total number of $Fe_{Tet}$ and $Fe_{Oct}$ has changed (see Table S1 in the Supplemental Material). If we apply formula (1), we get an $m_{tot}$ of 1112 $\mu_B$ for the reconstructed nanocube vs 1232 $\mu_B$ for the unreconstructed one. In order to validate the equation, we performed a series of HSE calculations, where we fully relaxed the nanocube atomic positions while varying $m_{tot}$ (see Fig. S1 in the Supplemental Material). The minimum total energy is registered for $m_{tot} = 1112$ $\mu_B$, in perfect agreement with the output by Eq. (1). Therefore, we conclude that this formula is rather general since it works for different situations ranging from bulk to surface to nanocube. The reconstruction reduces the $m_{tot}$ of the NP by about 10%.

Since in many biomedical applications magnetite NPs are spherical [30,40,41,42], we prepared another model of stoichiometric curved NPs [(Fe$_3$O$_4$)$_{136}$(H$_2$O)$_{18}$] by carving a sphere of 2.5 nm diameter and including 18 dissociatively adsorbed water molecules to saturate the too-low-coordinated Fe and O ions. Different from what was observed for nanocubes, for spherical NPs the N(Fe$_{Tet}^{2+}$) term is not null (as discussed below) and must be obtained by DFT calculations. Therefore, we performed a series of full atomic relaxation calculations with the HSE06 functional at different $m_{tot}$ values to determine the optimal $m_{tot}$ (600 $\mu_B$), as shown in Fig. S2 in the Supplemental Material. Using the Fe ions distribution in lowest energy configuration and formula (1), we get $m_{tot}$ of 602 $\mu_B$ with a 0.3% error with respect to the HSE optimal value. Considering the complexity of the curved surface, this simple formula works more than satisfactorily.

During the annealing process, simulated with DFTB+U, large atomic rearrangements occur at the curved surface to reduce the number of low-coordinated Fe ions (see Table S1 in the Supplemental Material), which involves the conversion of some six- and four-coordinated Fe$_{Oct}$ ions into five-coordinated ones and, in parallel, the conversion of some three-coordinated Fe$_{Tet}$ ions into four-coordinated ones. The resulting energy stabilization is of about 14 meV per atom. The annealed structure was then fully relaxed with the HSE functional (see Fig. 1). Similar atomic rearrangements are expected for larger spherical NPs. With the size increasing, the NP becomes more and more faceted and the percentage of low-coordinated cations decreases, as found in our previous work on TiO$_2$ NPs [43].

DFTB+U and HSE structures are compared by simulating the extended X-ray adsorption fine structure in real space (see Fig. S3 and Fig. S4 and detailed discussion in the Supplemental Material). The satisfactory agreement confirms the suitability of DFTB+U for the description of structural properties of magnetite NPs and supports its use for thermal annealing simulations.

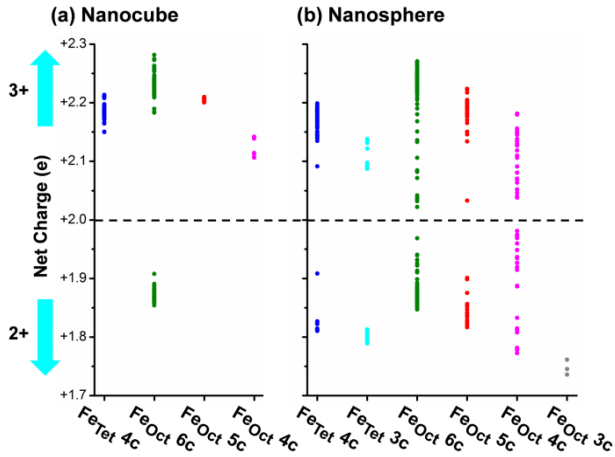

**Figure 3.** The net charge distribution for the Fe ions with different coordination at tetrahedral and octahedral sites in the optimized $Fe_3O_4$ (a) nanocube and (b) nanosphere with HSE functional after simulated annealing. Labels 3c–6c indicate the actual coordination number of the corresponding ions.

The charge ordering in magnetite is an interesting and challenging topic. Large efforts have been devoted to its understanding in the case of bulk during recent years [28,44,45,46,47,48,49], whereas, up to now, no information has been reported in the case of NPs. Hybrid functional calculations can provide precious information on charge distribution [28]. Here, we analyze Mulliken population charges, based on HSE calculations, to determine the $Fe^{2+}$ and $Fe^{3+}$ distribution. For the nanocube, we observe that all the $Fe_{Tet}$ ions are 3+, whereas $Fe_{Oct}$ are divided into two groups: $Fe_{Oct}^{2+}$ and $Fe_{Oct}^{3+}$, as shown in Fig. 3(a). All the low-coordinated $Fe_{Oct}$ ions (5c and 4c) on the NP surface are charged 3+. We show dissected views of the nanocube (Fig. 4, top) indicating an interesting core-shell structure with $Fe^{3+}$ ions in the outer-shell layers and alternating $Fe^{2+}/Fe^{3+}$ ions in the core.

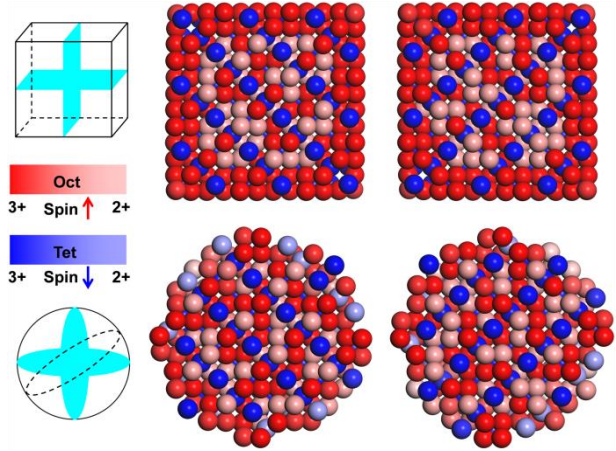

**Figure 4.** Selected dissected views showing the charge and spin distribution in the magnetite nanocube and nanosphere optimized models by HSE calculations. Oxygen atoms are not shown.

Different from the nanocube, the charge distribution on $Fe_{Oct}$ becomes blurred for the nanosphere [see Fig. 3(b)]. For discussion convenience, Fe ions with a net charge less than +2 are labeled as $Fe^{2+}$ and those with net charge more than +2 are labeled as $Fe^{3+}$. Interestingly, $Fe_{Tet}^{2+}$ ions, not present in bulk, surface or cubic NP systems, arise on the nanosphere surface because they are mostly three-coordinated and highly distorted (see light blue spheres in Fig. 4). $Fe_{Oct}^{2+}$ ions are distributed both in the core and at the surface of the NP (see light red spheres in Fig. 4).

**Table 1.** Average atomic magnetic moments ($m$) of Fe ions at tetrahedral and octahedral sites, total magnetic moment ($m_{tot}$) and saturation magnetization ($M_S$) for the magnetite nanocube and nanosphere (HSE) models.

|  | nanocube | Nanosphere |
|---|---|---|
| $m(Fe_{Tet}^{3+})$ ($\mu_B$) | -4.21 | -4.18 |
| $m(Fe_{Tet}^{2+})$ ($\mu_B$) | … | -3.70 |
| $m(Fe_{Oct}^{3+})$ ($\mu_B$) | 4.27 | 4.20 |
| $m(Fe_{Oct}^{2+})$ ($\mu_B$) | 3.75 | 3.78 |
| $m_{tot}$ ($\mu_B$) | 1112 | 600 |
| $M_S$ (emu/g) | 130.9 | 105.3 |

The atomic magnetic moments of $Fe^{2+}$ and $Fe^{3+}$ in the NPs (listed in Table 1) are similar to those calculated for the bulk corresponding species [28], except for $Fe_{Tet}^{2+}$.

The absolute value of the atomic magnetic moment of $Fe_{Tet}^{2+}$ is similar to that of $Fe_{Oct}^{2+}$.

We may note that, similar to bulk, in both nanocube and nanosphere models, $Fe_{Tet}$ ions always couple antiferromagnetically with $Fe_{Oct}$ ions. Being the latter in excess, the NPs are ferrimagnetic (Fig. 4 and Table 1).

As a further step in our analysis of the magnetic properties of magnetite nanoparticles, starting from the calculated optimal $m_{tot}$ value for the model NPs, presented in the previous section and reported also in Table 1, it is possible to estimate the $M_S$ value (per gram) for an ideal macroscopic sample of identical NPs. Such quantity can be compared with experimental measurements. For the nanocube, we obtain $M_S$ = 130.9 emu/g, which is larger than that for the nanosphere (105.3 emu/g), in agreement with experimental observations [15]. This result can be rationalized by analyzing the number of $Fe_{Oct}$ and $Fe_{Tet}$ ions ($N(Fe_{Oct})$ and $N(Fe_{Tet})$) in the NP. Because $Fe_{Oct}$ and $Fe_{Tet}$ couple antiferromagnetically, the net magnetic moment is determined by the excess of $Fe_{Oct}$, i.e. the difference between $N(Fe_{Oct})$ and $N(Fe_{Tet})$. The ratio of $N(Fe_{Oct})/N(Fe_{Tet})$ in the nanocube (2.3) is larger than that in the nanosphere (2.0) (see Table S1 in the SI), which accounts for the larger $m_{tot}$ and, thus, larger $M_S$ of the nanocube. From this point of view, nanocubes are more desirable for biomedical applications. We must note that the calculated $M_S$ values are larger than the experimental ones (54.0 emu/g for nanocubes with a size of 6.5 nm at 5 K [16] and about 35 emu/g for nanospheres with diameter of 5 nm at 20 K [24]). This is probably due to the antiphase boundary structural defects in the experimental NPs, which can largely reduce the magnetization [50]. In addition, the presence of nonmagnetic surfactants on the particle surface in experiments can also be a cause of magnetization reduction.

Both the nanocube and the nanosphere models possess larger $M_S$ than bulk (96 emu/g), which suggests that the larger the surface-to-bulk ratio, the larger the $M_S$. Therefore, the $M_S$ should decrease with the size increase. We calculated the $M_S$ of nanocubes of different sizes through formula (1) and the results confirm this trend, as shown in Fig. S5 in the Supplemental Material.

To get further insight into the electronic properties of magnetite NPs, we also present total and projected density of states on the *d* states of different Fe ions (Fig. S6) with the HSE method. NPs of both shapes have semiconducting character, which agrees with the tunneling microscopy measurements [25,26]. The nanocube possesses very similar electronic structure to that of bulk [28] and of the (001) surface [29], with a band gap of 0.55 eV. The conductivity is dominated by electron hopping between fully coordinated $Fe_{Oct}^{2+}$ and $Fe_{Oct}^{3+}$ in the core of the nanocube [Fig. S6(b)], but not on the surface because there only $Fe^{3+}$ ions are present (see Fig. 4, top). For the nanosphere, new surface states arise below the Fermi level (see Fig. S6 and S7 in the Supplemental Material). However, the band gap (0.53 eV) is similar to nanocubes. Electron hopping can take place both in the bulk and on the surface, thanks to the presence of $Fe_{Oct}^{2+}/Fe_{Oct}^{3+}$ and $Fe_{Tet}^{2+}/Fe_{Tet}^{3+}$ (see Fig. 4).

In summary, by adopting a set of proper methods, including a general empirical formula for the a-priori determination of the optimal total magnetic moment and the combination of DFTB and hybrid DFT, we have accomplished the QM simulation of cubic and spherical $Fe_3O_4$ NPs of realistic size (2.3-2.5 nm). The optimized atomic structures were obtained through simulated annealing at 500 K by MD with the DFTB method, followed by full atomic relaxation with hybrid DFT. From the $m_{tot}$ of one NP model, we can derive the macroscopic $M_S$ for an ideal sample of all identical NPs, to be compared with experimentally measured values.

Our results reveal the surface reconstruction mechanism that takes place at the four Fe-exposing corners of the nanocube and that reduces the $m_{tot}$ of the NP. Large atomic rearrangements also occur at the curved surface of the nanosphere to reduce the number of exposed low-coordinated Fe ions. The nanocube exhibits an interesting core-shell structure with respect to the distribution of $Fe^{2+}$ and $Fe^{3+}$, resulting in an insulating state in the shell and a semiconducting one in the core. In contrast, the appearance of $Fe_{Tet}^{2+}$ on the surface of the nanosphere makes the electrons hopping between $Fe_{Tet}^{2+}$ and $Fe_{Tet}^{3+}$ on the surface possible. Cubic NPs possess larger $M_S$ than spherical ones due to the larger ratio of $N(Fe_{Oct})/N(Fe_{Tet})$ and thus are more desirable for biomedical applications.

The approach of formula (1) for the a-priori determination of the optimal total magnetic moment, based on the principles of the crystal field theory, is also expected to be applicable to other magnetic materials of the spinel group, such as $MnFe_2O_4$, $NiFe_2O_4$, $CoFe_2O_4$, and so on. Therefore, our Letter not only makes possible to achieve the correct description, at the first-principles level of theory, of $Fe_3O_4$ nanoparticles, but also paves the way for the modeling of nanostructures of other similar magnetic materials.

**Acknowledgment**

The authors are grateful to Annabella Selloni and Gotthard Seifert for fruitful discussions and to Lorenzo Ferraro for his technical help. The project has received funding from the European Research Council (ERC) under the European Union's HORIZON2020 research and innovation programme (ERC Grant Agreement No [647020]).

**Additional information**

Supplemental Material is available …..

**References**

[1] J. M. Perez, L. Josephson, T. O'Loughlin, D. Högemann and R. Weissleder, Magnetic relaxation switches capable of sensing molecular interactions. Nat. Biotechnol. **20**, 816-820 (2002).

[2] J. Liu, Z. Sun, Y. Deng, Y. Zou, C. Li, X. Guo, L. Xiong, Y. Gao, F. Li, and D. Zhao, Highly Water-Dispersible Biocompatible Magnetite Particles with Low Cytotoxicity Stabilized by Citrate Groups. Angew. Chem. Int. Ed., **48**, 5875-5879 (2009).

[3] W. Wu, Z. Wu, T. Yu, C. Jiang and W. Kim, Recent progress on magnetic iron

oxide nanoparticles: synthesis, surface functional strategies and biomedical applications. Sci. Technol. Adv. Mater. **16**, 023501 (2015).

[4] Q. A. Pankhurst, N. T. K. Thanh, S. K. Jones and J. Dobson, Progress in applications of magnetic nanoparticles in biomedicine. J. Phys. D: Appl. Phys. **42**, 224001 (2009).

[5] A. K. Gupta and M. Gupta, Synthesis and surface engineering of iron oxide nanoparticles for biomedical applications. Biomaterials **26**, 3995-4021 (2005).

[6] C. Sun, J. S. Lee and M. Zhang, Magnetic nanoparticles in MR imaging and drug delivery. Adv. Drug Deliv. Rev. **60**, 1252-1265 (2008).

[7] Q. A. Pankhurst, J. Connolly, S. K. Jones and J. Dobson, Applications of magnetic nanoparticles in biomedicine. J. Phys. D: Appl. Phys. **36**, R167-R181 (2003).

[8] S. Laurent, D. Forge, M. Port, A. Roch, C. Robic, L. V. Elst and R. N. Muller, Magnetic Iron Oxide Nanoparticles: Synthesis, Stabilization, Vectorization, Physicochemical Characterizations, and Biological Applications. Chem. Rev. **108**, 2064-2110 (2008).

[9] M. Colombo, S. Carregal-Romero, M. F. Casula, L. Gutiérrez, M. P. Morales, I. B. Böhm, J. T. Heverhagen, D. Prosperi and W. J. Parak, Biological applications of magnetic nanoparticles. Chem. Soc. Rev. **41**, 4306-4334 (2012).

[10] S. Sun and H. Zeng, Size-Controlled Synthesis of Magnetite Nanoparticles. J. Am. Chem. Soc. **124**, 8204-8205 (2002).

[11] Y. Hou, J. Yu and S. Gao, Solvothermal reduction synthesis and characterization of superparamagnetic magnetite nanoparticles. J. Mater. Chem. **13**, 1983-1987 (2003).

[12] S. Sun, H. Zeng, D. B. Robinson, S. Raoux, P. M. Rice, S. X. Wang and G. Li, Monodisperse $MFe_2O_4$ (M = Fe, Co, Mn) Nanoparticles. J. Am. Chem. Soc. **126**, 273-279 (2004).

[13] J. Park, K. An, Y. Hwang, J. Park, H. Noh, J. Kim, J. Park, N. Hwang and T. Hyeon, Ultra-large-scale syntheses of monodisperse nanocrystals. Nat. Mater. **3**, 891-895 (2004).

[14] Y. Tian, B. Yu, X. Li and K. Li, Facile solvothermal synthesis of monodisperse $Fe_3O_4$ nanocrystals with precise size control of one nanometre as potential MRI contrast agents. J. Mater. Chem. **21**, 2476 (2011).

[15] M. V. Kovalenko, M. I. Bodnarchuk, R. T. Lechner, G. Hesser, F. Schäffler and W. Heiss, Fatty Acid Salts as Stabilizers in Size- and Shape-Controlled Nanocrystal Synthesis: The Case of Inverse Spinel Iron Oxide. J. Am. Chem. Soc. **129**, 6352-6353 (2007).

[16] H.Yang, T. Ogawa, D. Hasegawa and M. Takahashi, Synthesis and magnetic properties of monodisperse magnetite nanocubes. J. Appl. Phys. **103**, 07D526 (2008).

[17] D. Kim, N. Lee, M. Park, B. H. Kim, K. An and T. Hyeon, Synthesis of Uniform Ferrimagnetic Magnetite Nanocubes. J. Am. Chem. Soc. **131**, 454–455 (2009).

[18] L. Zhao and L. Duan, Uniform $Fe_3O_4$ Octahedra with Tunable Edge Length-Synthesis by a Facile Polyol Route and Magnetic Properties. Eur. J. Inorg. Chem. 5635–5639 (2010).

[19] L. Zhang, J. Wu, H. Liao, Y. Hou and S. Gao, Octahedral $Fe_3O_4$ nanoparticles and their assembled structures. Chem. Commun. 4378–4380 (2009).

[20] X. Li, D. Liu, S. Song, X. Wang, X. Ge and H. Zhang, Rhombic dodecahedral $Fe_3O_4$: ionic liquid-modulated and microwave-assisted synthesis and their magnetic properties. CrystEngComm, **13**, 6017–6020 (2011).

[21] X. Cheng, J. Jiang, D. Jiang and Z. Zhao, Synthesis of Rhombic Dodecahedral $Fe_3O_4$ Nanocrystals with Exposed High-Energy {110} Facets and Their Peroxidase-like Activity and Lithium Storage Properties. J. Phys. Chem. C **118**, 12588−12598 (2014).

[22] R. Zheng, H. Gu, B. Xu, K. K. Fung, X. Zhang and S. P. Ringer, Self-Assembly and Self-Orientation of Truncated Octahedral Magnetite Nanocrystals. Adv. Mater. **18**, 2418–2421 (2006).

[23] L. Zhao, H. Zhang, Y. Xing, S. Song, S. Yu, W. Shi, X. Guo, J. Yang, Y. Lei and F. Cao, Morphology-Controlled Synthesis of Magnetites with Nanoporous Structures and Excellent Magnetic Properties. Chem. Mater. **20**, 198-204 (2008).

[24] K. Woo, J. Hong, S. Choi, H. Lee, J. Ahn, C. S. Kim and S. W. Lee, Easy Synthesis and Magnetic Properties of Iron Oxide Nanoparticles. Chem. Mater. **16**, 2814-2818 (2004).

[25] A. Hevroni, M. Bapna, S. Piotrowski, S. A. Majetich and G. Markovich, Tracking the Verwey Transition in Single Magnetite Nanocrystals by Variable-Temperature Scanning Tunneling Microscopy. J. Phys. Chem. Lett. **7**, 1661−1666 (2016).

[26] Q. Yu, Verwey transition in single magnetite nanoparticles. Phys. Rev. B **90**,

075122 (2014).

[27] R. Bliem, E. McDermott, P. Ferstl, M. Setvin, O. Gamba, J. Pavelec, M. A. Schneider, M. Schmid, U. Diebold, P. Blaha, L. Hammer and G. S. Parkinson, Subsurface cation vacancy stabilization of the magnetite (001) surface. Science **346**, 1215 (2014).

[28] H. Liu and C. Di Valentin, Band Gap in Magnetite above Verwey Temperature Induced by Symmetry Breaking. J. Phys. Chem. C **121**, 25736 (2017).

[29] H. Liu and C. Di Valentin, Bulk-terminated or reconstructed $Fe_3O_4$(001) surface: water makes a difference. Nanoscale **10**, 11021 (2018).

[30] M. Patitsa, K. Karathanou, Z. Kanaki, L. Tzioga, N. Pippa, C. Demetzos, D. A. Verganelakis, Z. Cournia and A. Klinakis, Magnetic nanoparticles coated with polyarabic acid demonstrate enhanced drug delivery and imaging properties for cancer theranostic applications. Sci. Rep. **7**, 775 (2017).

[31] R.A. Harris, H. van der Walt and P.M. Shumbula, Molecular dynamics study on iron oxide nanoparticles stabilised with Sebacic Acid and 1,10-Decanediol surfactants. J. Mol. Struct. **1048**, 18-26 (2013).

[32] R. A. Harris, P. M. Shumbula and H. van der Walt, Analysis of the Interaction of Surfactants Oleic Acid and Oleylamine with Iron Oxide Nanoparticles through Molecular Mechanics Modeling Langmuir **31**, 3934-3943 (2015).

[33] A. Hosseini nasr, H. Akbarzadeh and R. Tayebee, Adsorption mechanism of different acyclovir concentrations on 1–2 nm sized magnetite nanoparticles: A molecular dynamics study J. Mol. Liq. **254**, 64-69 (2018).

[34] M. Elstner, D. Porezag, G. Jungnickel, J. Elsner, M. Haugk, Th. Frauenheim, S. Suhai and G. Seifert, Self-consistent-charge density-functional tight-binding method for simulations of complex materials properties. Phys. Rev. B **58**, 7260 (1998).

[35] A. V. Krukau, O. A. Vydrov, A. F. Izmaylov and G. E. Scuseria, Influence of the exchange screening parameter on the performance of screened hybrid functionals. *J. Chem. Phys.* **125**, 224106 (2006).

[36] G. Zheng, H. A. Witek, P. Bobadova-Parvanova, S. Irle, D. G. Musaev, R. Prabhakar and K. Morokuma, Parameter Calibration of Transition-Metal Elements for the Spin-Polarized Self-Consistent-Charge Density-Functional Tight-Binding (DFTB) Method: Sc, Ti, Fe, Co, and Ni. J. Chem. Theory Comput. **3**, 1349 (2007).
15

[37] R. Dovesi, R. Orlando, A. Erba, C. M. Zicovich-Wilson, B. Civalleri, S. Casassa, L. Maschio, M. Ferrabone, M. De La Pierre, P. D'Arco, et al. CRYSTAL14: A program for the ab initio investigation of crystalline solids. *Int. J. Quantum Chem.*, **114**, 1287-1317 (2014).

[38] R. Dovesi, V. R. Saunders, C. Roetti, R. Orlando, C. M. Zicovich-Wilson, F. Pascale, B. Civalleri, K. Doll, N. M. Harrison, I. J. Bush, et al. Crystal14 User's Manual. University of Torino: Torino, Italy, **2014**.

[39] H. Liu, G. Seifert and C. Di Valentin, An efficient way to model complex magnetite: assessment of SCC-DFTB against DFT. J. Chem. Phys. **150**, 094703 (2019).

[40] Y. Jun, et al. Nanoscale Size Effect of Magnetic Nanocrystals and Their Utilization for Cancer Diagnosis via Magnetic Resonance Imaging. J. Am. Chem. Soc. **127**, 5732-5733 (2005).

[41] D. Ling, et al. Multifunctional Tumor pH-Sensitive Self-Assembled Nanoparticles for Bimodal Imaging and Treatment of Resistant Heterogeneous Tumors. J. Am. Chem. Soc. **136**, 5647−5655 (2014).

[42] C. Prashant, Superparamagnetic iron oxide e Loaded poly (lactic acid)–D-a-tocopherol polyethylene glycol 1000 succinate copolymer nanoparticles as MRI contrast agent. Biomaterials **31**, 5588-5597 (2010).

[43] Selli, D., Fazio, G. & Di Valentin C. Modelling realistic $TiO_2$ nanospheres: a benchmark study of SCC-DFTB against hybrid DFT. J. Chem. Phys. **147**, 164701 (2017).

[44] J. P. Wright, J. P. Attfield and P. G. Radaelli, Long Range Charge Ordering in Magnetite Below the Verwey Transition. Phys. Rev. Lett. **87**, 266401 (2001).

[45] I. Leonov, A. N. Yaresko, V. N. Antonov, M. A. Korotin and V. I. Anisimov, Charge and Orbital Order in $Fe_3O_4$. Phys. Rev. Lett. **93**, 146404 (2004).

[46] H. Jeng, G.Y. Guo and D. J. Huang, Charge-Orbital Ordering and Verwey Transition in Magnetite. Phys. Rev. Lett. **93**, 156403 (2004).

[47] J. Schlappa, C. Schüßler-Langeheine, C. F. Chang, H. Ott, A. Tanaka, Z. Hu, M.W. Haverkort, E. Schierle, E. Weschke, G. Kaindl and L. H. Tjeng, Direct Observation of $t_{2g}$ Orbital Ordering in Magnetite. Phys. Rev. Lett. **100**, 026406 (2008).

[48] S. Weng, Y. Lee, C. Chen, C. Chu, Y. Soo and S. Chang, Direct Observation of

Charge Ordering in Magnetite Using Resonant Multiwave X-Ray Diffraction. Phys. Rev. Lett. **108**, 146404 (2012).

[49] M. S. Senn, J. P. Wright and J. P. Attfield, Charge order and three-site distortions in the Verwey structure of magnetite. Nature, **481**, 171-176 (2012).

[50] Z. Nedelkoski, D. Kepaptsoglou, L. Lari, T. Wen, R. A. Booth, et al. Origin of reduced magnetization and domain formation in small magnetite nanoparticles. Sci. Rep. **7**, 45997 (2017).

**Supplemental Material**

**Shaping Magnetite Nanoparticles from First-principles**

Hongsheng Liu, Cristiana Di Valentin*

*Dipartimento di Scienza dei Materiali, Università di Milano-Bicocca, via R. Cozzi 55, I-20125 Milano, Italy*

**Computational details**

*Magnetic moment definition*

To properly describe magnetite, high-level quantum mechanical simulations beyond standard density functional theory (DFT), such as DFT+U and hybrid DFT are necessary [1,2]. Up to now, two kinds of basis sets for DFT calculations are available, i.e. plain wave and Gaussian-type atomic orbital basis sets. In the case of periodic calculations with a plain wave basis set, the total magnetic moment of magnetite NPs can be relaxed during atomic optimization. However, a large surrounding vacuum is required to model nanoparticles with a periodic approach, which demands a very large memory and makes the calculation too expensive to be performed. In the case of molecular calculations with an atomic orbital basis set, there is no need for a surrounding vacuum, making the simulation of large magnetite NPs (1000-1400 atoms) become possible though still very expensive. However, with the latter approach the total magnetic moment cannot be freely relaxed because the nanoparticle is treated as a molecular model, whose spin multiplicity must be explicitly given. This is the reason why equation (1) in the manuscript is extremely relevant and useful.

*DFT calculations*

Hybrid functional calculations (HSE06 [3]) were carried out using the CRYSTAL17 package [4,5] based on DFT where the Kohn−Sham orbitals are expanded in Gaussian-type orbitals (the all-electron basis sets are H|5-11G*, O|8-411G* and Fe|8-6-411G*, according to the scheme previously used for $Fe_3O_4$ [1]). The convergence criterion of 0.02 eV/Å for force was used during geometry optimization and the convergence criterion for total energy was set at $10^{-6}$ Hartree for all the calculations.

*DFTB calculations*

Even with the atomic orbital basis set, the molecular dynamics calculations for $Fe_3O_4$ NPs, which is important for searching the global minimum structure, are still too expensive to be performed. Cheaper but effective methods are needed. Self-consistent charge density functional tight binding (SCC-DFTB) is recently proved to be an effective method for the description of magnetite bulk and surfaces [6]. The SCC-DFTB method is an approximated DFT-based method that derives from the second-order expansion of the Kohn-Sham total energy in DFT with respect to the electron density fluctuations. The SCC-DFTB total energy can be defined as:

$$E_{tot} = \sum_i^{occ} \varepsilon_i + \frac{1}{2}\sum_{\alpha,\beta}^N \gamma_{\alpha\beta}\Delta q_\alpha \Delta q_\beta + E_{rep} \qquad (2),$$

where, the first term is the sum of the one-electron energies $\varepsilon_i$ coming from the diagonalization of an approximated Hamiltonian matrix. $\Delta q_\alpha$ and $\Delta q_\beta$ are the induced charges on the atoms α and β, respectively, and $\gamma_{\alpha\beta}$ is a Coulombic-like interaction potential. $E_{rep}$ is a short-range pairwise repulsive potential. More details about the SCC-DFTB method can be found in Refs.7, 8, and 9. DFTB will be used as shorthand for SCC-DFTB.

The DFTB calculations were performed by the DFTB+ package [10] without imposing any symmetry constraint. For the Fe-Fe and Fe-H interactions we used the "trans3d-0-1" set of parameters as reported previously [11]. For the O-O, H-O and H-H interactions we used the "mio-1-1" set of parameters [7]. For the Fe-O interactions, we used the Slater-Koster files fitted by us previously, which can well reproduce the results for magnetite bulk and surfaces from HSE06 and PBE+U calculations [6]. To properly deal with the strong correlation effects among Fe 3d electrons, DFTB+U [12] with an effective U-J value of 3.5 eV was adopted according to our previous work on magnetite bulk and (001) surface [1,2]. The convergence criterion of $10^{-4}$ a.u. for force was used during geometry optimization and the convergence threshold on the self-consistent charge (SCC) procedure was set to be $10^{-5}$ a.u.

DFTB molecular dynamics was performed within the canonical ensemble (NVT) with a time step of 1 fs. An Andersen thermostat [13] was used to target the desired temperatures. To simulate the temperature annealing processes, the nanocube and nanosphere were quickly heated up to 500 K (within 0.5 ps) and then kept at 500 K for 3.5 ps, and then cooled down slowly to 100 K when no further structural changes were expected. The total simulation time for the nanocube and nanosphere are 16 ps and 22 ps, respectively.

**Table S1.** Numbers of Fe ions at tetrahedral ($Fe_{Tet}$) and octahedral ($Fe_{Oct}$) sites with different coordination (3c, 4c, 5c and 6c) in the carved and optimized (relaxed with HSE after annealing) $Fe_3O_4$ nanoparticles.

|  | Nanocube | | Nanosphere | |
| --- | --- | --- | --- | --- |
|  | carved | optimized | carved | optimized |
| $Fe_{Tet}$ 4c | 170 | 182 | 76 | 96 |
| $Fe_{Tet}$ 3c | 0 | 0 | 60 | 40 |
| $Fe_{Oct}$ 6c | 252 | 240 | 170 | 161 |
| $Fe_{Oct}$ 5c | 144 | 144 | 42 | 55 |
| $Fe_{Oct}$ 4c | 36 | 36 | 60 | 53 |
| $Fe_{Oct}$ 3c | 0 | 0 | 0 | 3 |

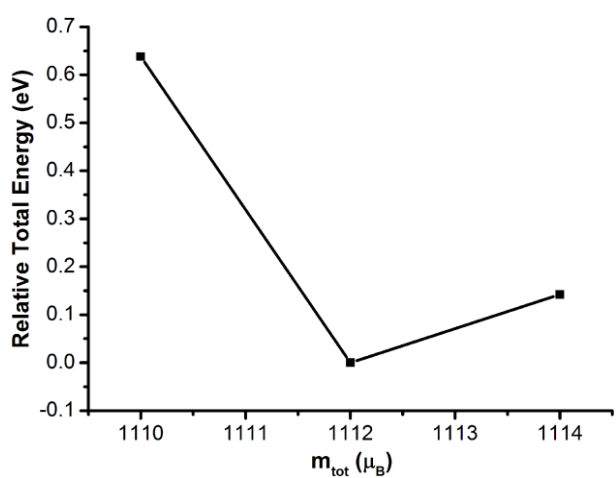

**Figure S1.** The relative total energy as a function of the total magnetic moment ($m_{tot}$) for the nanocube after annealing. With each $m_{tot}$ the nanocube is fully relaxed with HSE method.

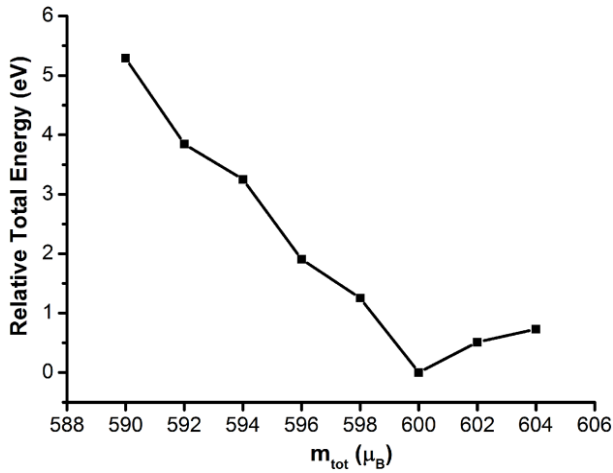

**Figure S2.** The relative total energy as a function of the total magnetic moment ($m_{tot}$) for the nanosphere before annealing. With each $m_{tot}$ the nanosphere is fully relaxed with HSE method with 40% of exact exchange. This is because before annealing, the calculation with HSE (25% of exact exchange) could not converge. For bulk magnetite, HSE with 40% of exact exchange gives the same optimal total magnetic moment as that given by HSE with 25% of exact exchange. Therefore, HSE with 40% of exact exchange should be acceptable for the searching of the optimal total magnetic moment.

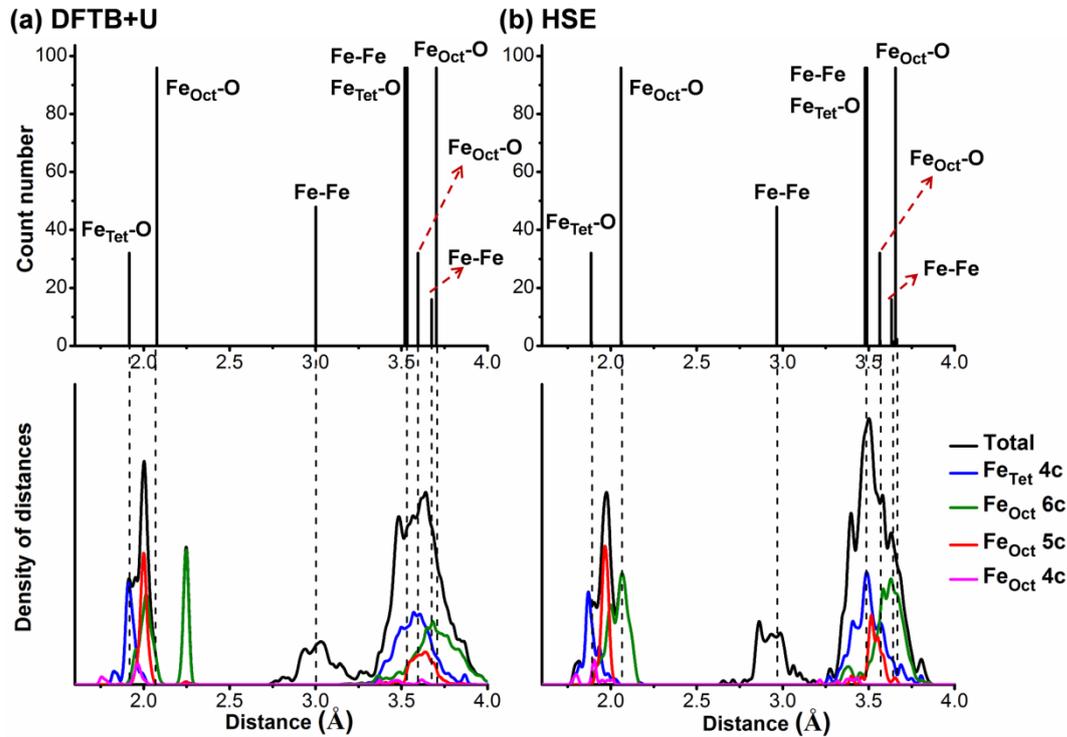

**Figure S3.** Distances distribution (EXAFS in real space) computed with DFTB+U and HSE for magnetite bulk (top panel) and nanocube after annealing (bottom panel).

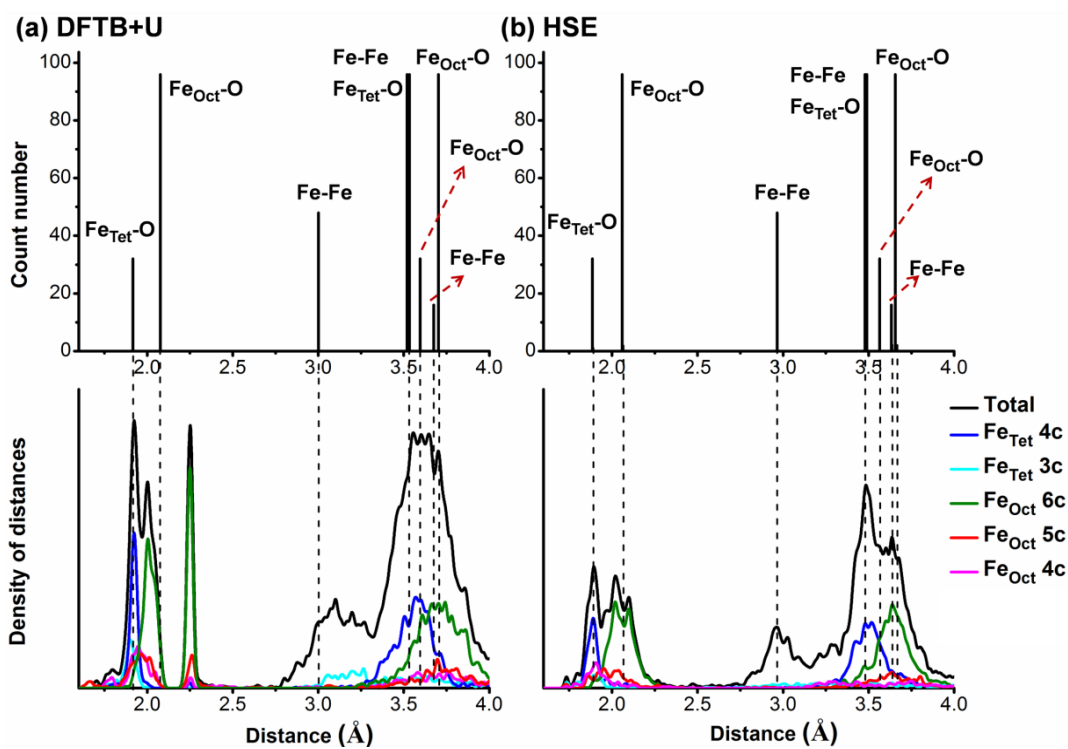

**Figure S4.** Distances distribution (EXAFS in real space) computed with DFTB+U and HSE for magnetite bulk (top panel) and the nanosphere after annealing (bottom panel).

The extended X-ray adsorption fine structure (EXAFS) in real space was simulated for magnetite bulk and nanoparticles (Figure S3 and S4) to analyze the structural distortion and compare the structures relaxed with HSE06 and DFTB+U. The real space EXAFS was simulated by calculating the density of distances for each Fe ions with other (Fe or O) ions and projecting them on Fe ions with different coordination. In general, the $Fe_{Tet}$-O and $Fe_{Oct}$-O bond lengths in bulk magnetite are broadened in nanoparticles. The peak for the $Fe_{Oct}$-O bond length in the bulk splits into several peaks for $Fe_{Oct}$ with different coordination, the lower the coordination the shorter the bond length. For both the nanocube and the nanosphere, satisfactory agreement between DFTB+U and HSE can be found with respect to the position of peaks for $Fe_{Tet}$-O and $Fe_{Oct}$-O distances, which indicates the good performance of DFTB+U on magnetite nanoparticles.

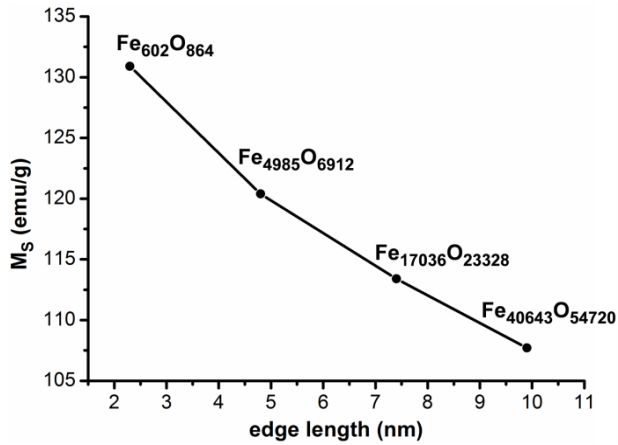

**Figure S5.** The saturation magnetization of nanocubes as a function of the size. The structures of large nanocubes ($Fe_{4985}O_{6912}$, $Fe_{17036}O_{23328}$ and $Fe_{40643}O_{54720}$) are extrapolated from the small one ($Fe_{602}O_{864}$). The saturation magnetization is calculated through formula (1) in the manuscript.

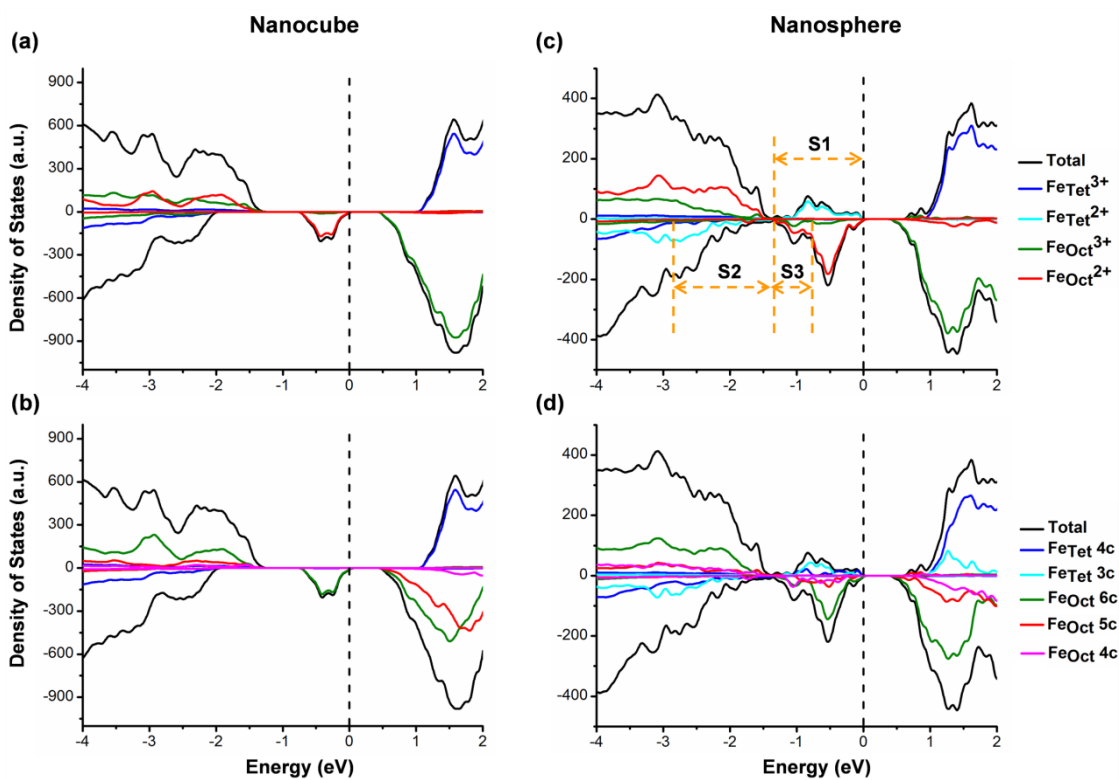

**Figure S6.** PDOS on the d states of different Fe ions in the nanocube, (a)-(b), and nanosphere, (c)-(d), optimized models by HSE method, after simulated annealing. In (a) and (c), the projection is on Fe ions at tetrahedral and octahedral sites with different atomic charge, whereas in (b) and (d), it is on Fe ions at tetrahedral and octahedral sites with different coordination number (3c, 4c, 5c and 6c). Legend of colors is on the right. The Fermi level is scaled to zero as indicated by the dashed black lines. S1, S2 and S3 in (c) are states in the nanosphere that are missing in the nanocube. The state S1 and S2 are from the d states of $Fe_{Tet}^{2+}$ ions, which are mainly 3-coordinated. The state S3 is from the d states of both $Fe_{Oct}^{2+}$ and $Fe_{Oct}^{3+}$, which are either 4-coordinated or 6-coordinated. All these new states are surface states, according to the charge density plots (Figure S7).

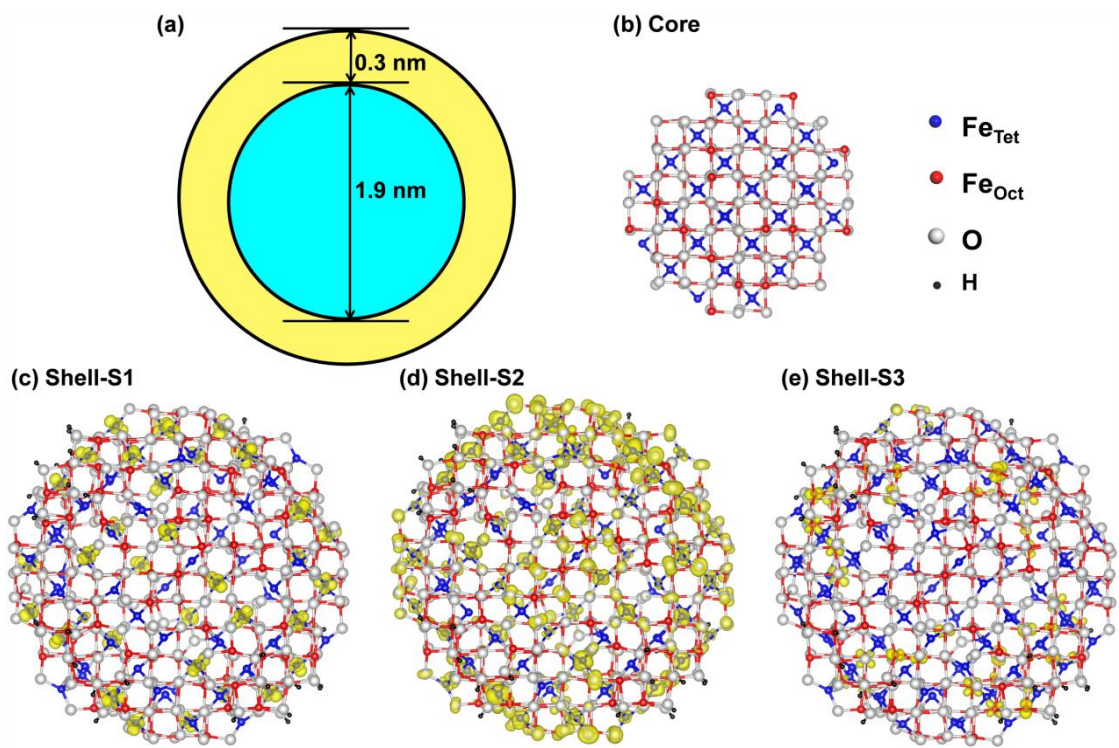

**Figure S7.** (a) Schematic diagram of the nanosphere with the core part (colored in cyan) and the shell part (colored in yellow). (b) to (d) are the charge density plots calculated with HSE for states S1, S2 and S3 marked in Figure S5. (b) is the plot on the core part of the nanosphere, which is the same for state S1, S2 and S3. (c) to (d) are the plots on the shell part of the nanosphere for state S1, S2 and S3, respectively. The isosurface level is 0.015 electron/bohr$^3$. These plots indicate that S1, S2 and S3 are all surface states.

**References:**

[1] H. Liu and C. Di Valentin, Band Gap in Magnetite above Verwey Temperature Induced by Symmetry Breaking. *J. Phys. Chem. C* **121**, 25736 (2017).

[2] H. Liu and C. Di Valentin, Bulk-terminated or reconstructed $Fe_3O_4$(001) surface: water makes a difference. *Nanoscale* **10**, 11021 (2018).

[3] A. V. Krukau, O. A. Vydrov, A. F. Izmaylov and G. E. Scuseria, Influence of the exchange screening parameter on the performance of screened hybrid functionals. *J. Chem. Phys.*, **125**, 224106 (2006).

[4] R. Dovesi, R. Orlando, A. Erba, C. M. Zicovich-Wilson, B. Civalleri, S. Casassa, L. Maschio, M. Ferrabone, M. De La Pierre, P. D'Arco, et al. CRYSTAL14: A program for the ab initio investigation of crystalline solids. *Int. J. Quantum Chem.*, **114**, 1287-1317 (2014).

[5] R. Dovesi, V. R. Saunders, C. Roetti, R. Orlando, C. M. Zicovich-Wilson, F.

Pascale, B. Civalleri, K. Doll, N. M. Harrison, I. J. Bush, et al. Crystal14 User's Manual. University of Torino: Torino, Italy, **2014**.

[6] H. Liu, G. Seifert and C. Di Valentin, An efficient way to model complex magnetite: assessment of SCC-DFTB against DFT. J. Chem. Phys. **150**, 094703 (2019).

[7] M. Elstner, D. Porezag, G. Jungnickel, J. Elsner, M. Haugk, Th. Frauenheim, S. Suhai and G. Seifert, Self-consistent-charge density-functional tight-binding method for simulations of complex materials properties. Phys. Rev. B **58**, 7260 (1998).

[8] G. Seifert and J. Joswig, Density-functional tight binding—an approximate density-functional theory method. Wiley Interdiscip. Rev.: Comput. Mol. Sci. **2**, 456 (2012).

[9] M. Elstner and G. Seifert, Density functional tight binding. Philos. Trans. R. Soc. A **372**, 20120483 (2014).

[10] B. Aradi, B. Hourahine and T. Frauenheim, DFTB+, a Sparse Matrix-Based Implementation of the DFTB Method. J. Phys. Chem. A **111**, 5678 (2007).

[11] G. Zheng, H. A. Witek, P. Bobadova-Parvanova, S. Irle, D. G. Musaev, R. Prabhakar and K. Morokuma, Parameter Calibration of Transition-Metal Elements for the Spin-Polarized Self-Consistent-Charge Density-Functional Tight-Binding (DFTB) Method: Sc, Ti, Fe, Co, and Ni. J. Chem. Theory Comput. **3**, 1349 (2007).

[12] B. Hourahine, S. Sanna, B. Aradi, C. Köhler, Th. Niehaus and Th. Frauenheim, Self-Interaction and Strong Correlation in DFTB. J. Phys. Chem. A **111**, 5671 (2007).

[13] H. C. Andersen, Molecular dynamics at constant pressure and/or temperature. J. Chem. Phys. **72**, 2384 (1980).